\documentclass[10pt]{iopart}
\makeatletter
\@namedef{ver@amsmath.sty}{}
\makeatother
\newcommand{\DeclareMathOperator}[2]{\newcommand{#1}{\operatorname{#2}}}
\usepackage{siunitx}
\usepackage{graphicx}
\usepackage{iopams}
\usepackage{physics}
\usepackage{color}
\usepackage{bm}
\usepackage{circuitikz}
\graphicspath{{figs/}}
\begin{document}
\newcommand{\vend}{v_\mathrm{end}}
\newcommand{\vfall}{v_\mathrm{fall}}
\newcommand{\iarc}{i_\mathrm{d}}
\newcommand{\tz}{t_\mathrm{0}}
\newcommand{\tend}{t_\mathrm{end}}
\newcommand{\uarc}{u_\mathrm{arc}}
\newcommand{\thv}{\bm{\mathrm{\theta}}}
\newcommand{\thvh}{\bm{\hat{\mathrm{\theta}}}}
\newcommand{\rs}[1]{_\mathrm{#1}}

\title[]{Numerical investigation of transient, low-power metal vapour discharges occurring in near limit ignitions of flammable gas}
\author{Rajiv Shekhar$^{1,2}$, Sergey Gortschakow$^3$, Holger Grosshans$^1$, Udo Gerlach$^1$, Dirk Uhrlandt$^3$}
\address{$^1$Physikalisch-Technische Bundesanstalt (PTB), Bundesallee 100, 38116 Braunschweig, Germany}
\address{$^2$School of ITEE, The University of Queensland, Qld, Australia}
\address{$^3$Leibniz Institute for Plasma Science and Technology (INP), 17489 Greifswald, Germany}
\ead{rajiv.shekhar@uqconnect.edu.au}
\vspace{10pt}
\begin{indented}
\item[]August 2017
\end{indented}

\begin{abstract}
This article presents an investigation of a transient (\SI{30}{\micro s} -- \SI{5}{ms}) electrical discharge in metal vapour with low voltage ($\le$ 30 V) and current ($\le$ 200 mA), drawn between two separating electrodes. Discharges of this type are rarely studied, but are important in electrical explosion safety, as they can ignite flammable gasses. An empirical model is developed based on transient recordings of discharge voltages and currents and high speed broadband image data. The model is used for predicting the electrical waveforms and spatial power distribution of the discharge. The predicted electrical waveforms show good accuracy under various scenarios. To further investigate the underlying physics, the model is then incorporated into a simplified 3-D gas dynamics simulation including molecular diffusion, heat transfer and evaporation of metal from the electrode surface. The local thermodynamic equilibrium (LTE) assumption is next used to calculate electrical conductivity from the simulated temperature fields, which in turn is integrated to produce electrical resistance over time. This resistance is then compared to that implied by the voltage and current waveforms predicted by the empirical model. The comparison shows a significant discrepancy, yielding the important insight that the studied discharge very likely deviates strongly from LTE.
\end{abstract}

%
%
\submitto{\JPD}
%
%
\ioptwocol

\section{Introduction}
\label{sec:introduction}
Electrical discharges have been studied extensively as a means of igniting flammable gasses in the field of combustion science. Previous works have typically focussed on high voltage spark discharges, possessing substantially different characteristics to the type of discharge considered here. Although scientific literature on the subject is limited, these low power discharges, referred to variously as break arcs, break sparks, or contact arcs, are well known in electrical explosion safety. Industry standards prescribe methods for managing the explosion risk they pose~\cite{iec2011,iec2011a}, but require experiment based methods for establishing safety limits which have been proven to be unreliable~\cite{klausmeyer_introduction_2014}. There is thus an increasing demand for a reliable alternative based on sound science to be developed, which requires a fundamental understanding of the relevant phenomena.

\begin{figure*}
\includegraphics{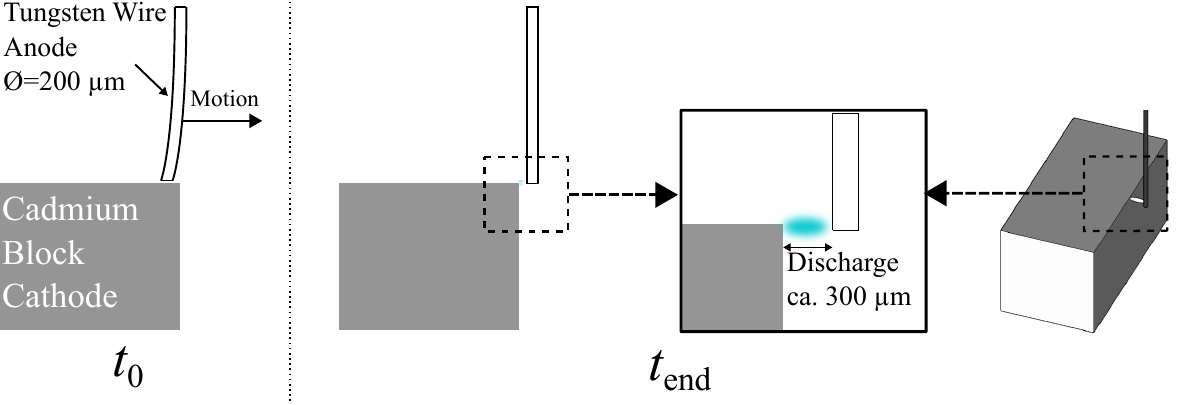}
\caption{Geometry of electrodes and discharge}
\label{fig:contact_arc}
\end{figure*}

The specific scenario being investigated is described in figure~\ref{fig:contact_arc}. A tungsten wire anode is positioned with its tip in contact with the surface of a cadmium block cathode. The wire moves along the surface until it reaches the edge of the block, where contact is broken. As the two electrodes separate, the discharge is drawn between them, reaching a typical maximum length of around \SI{300}{\micro m}. The electrode materials and geometry are based on a reference scenario for standardised explosion safety testing~\cite{iec2011}. Additionally, the electrodes are energised by an electronic circuit which adjusts output voltage so that a programmed constant current flow is maintained.

A previous investigation applied a combination of optical diagnostic experiments and reactive computational fluid dynamics simulation to the problem, focussing on the combustion resulting from the discharge~\cite{shekhar_ignition_2017}. Here, the discharge was represented by an energy source term derived from an empirical model of the discharge. The empirical model was based on a quasi-static correlation between voltage, current and length of the discharge. This work presents an expanded derivation of this empirical model, including several improvements and the incorporation of new experimental data. A partial validation is also presented utilising electrical measurement data. A key challenge of the empirical approach is quantifying how power is spatially distributed within the discharge. This is currently done using high speed broadband image data, which is difficult to analyse. Additionally, this aspect of the model is not easy to validate.

The direct imposition of a spatial power distribution could be avoided if electrical conductivity within the discharge region were known. This calculation would be significantly simplified by the assumption of local thermodynamic equilibrium (LTE), i.e.: that all species present at any point in the plasma have kinetic energies described by a Maxwell-Boltzmann distribution for a given single temperature. This has the important implication that the parameters of the plasma such as ionisation fraction and electrical conductivity are defined uniquely by this temperature. Previous studies of high voltage sparks conducted simulations under the LTE assumption, where tabulated data for plasma parameters as functions of temperature were used to couple the equations of gas dynamics to those of electromagnetics~\cite{ekici_thermal_2007,thiele_geometrical_2000}.

Applying a similar approach to modelling the low power discharge in question would require some indication of whether the LTE assumption is valid. In models of discharges in switching electrodes at high currents, the bulk of the plasma is often considered to be in LTE~\cite{karetta_simulation_1998}. Quasi-static empirical models are also used for equilibrium arcs, and have shown some success in this application~\cite{shekhar_modelling_2015}. The radiation emitted by the discharge was also confirmed to comprise predominantly of atomic lines of cadmium~\cite{uber_experimental2_2017}.

Characterising this discharge as an equilibrium arc would, however, be questionable, considering the extremely small energy and time scales involved. This work therefore additionally aims to provide insight into this matter. Given the substantial difficulties involved in modelling a non-equilibrium plasma, it is desirable to know beforehand whether doing so is necessary. For this purpose, the developed empirical model is used to describe the energy input of the discharge into a volume of gas. This model reflects measurements of the discharge and does not make assumptions with regard to the underlying physics. Next, assuming LTE, fields for temperature and electrical conductivity are obtained from the simulation. The latter is then integrated spatially to calculate electrical resistance, which can be compared to that predicted directly by the empirical source model. The consistency (or otherwise) between these two calculations provides an indication of the LTE assumption's validity.

The article is presented as follows: section \ref{sec:source} describes the development of the empirical discharge model, section \ref{sec:physics} describes the simulated solid, gas phase and LTE plasma physics, and section \ref{sec:results} presents simulation results and comparisons with the empirical model. Conclusions from the work and avenues for future development are presented in sections \ref{sec:discussion} and \ref{sec:conclusion}.

\section{Empirical discharge model}
\label{sec:source}
The empirical part of the model defines the total power input into the discharge over its duration, as well as how that power is distributed over the discharge region. This section outlines the derivation of these two relationships. A similar approach to Ref.~\cite{shekhar_ignition_2017,shekhar_modelling_2015} is taken, applying it to the recent dataset of Ref.~\cite{uber_experimental_2017} and adding a more detailed consideration of varying arc geometry. The dataset consists of voltage and current waveforms, together with corresponding high speed image recordings from which length variation of the discharge over time is measured. The data is thus a series of voltage/current/length triples -- one for each recorded time instant of each discharge.

\begin{figure*}
\includegraphics{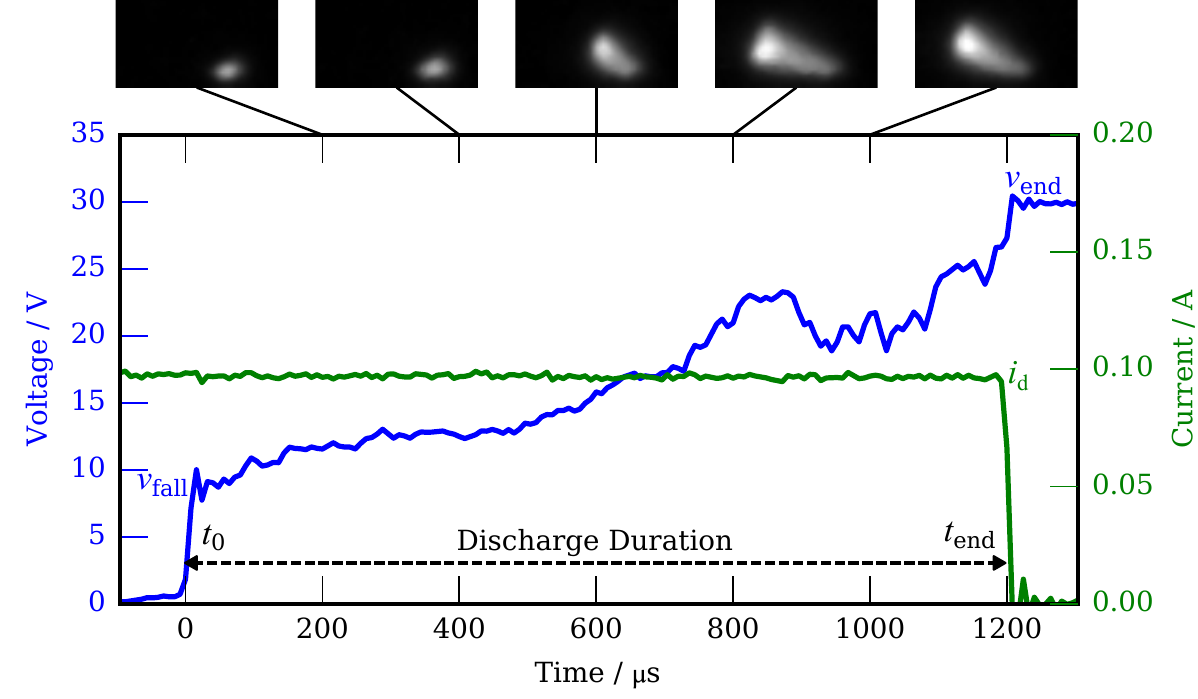}
\caption{Example of data recorded from one discharge}
\label{fig:data_example}
\end{figure*}

An example of this data for one discharge event is shown in figure \ref{fig:data_example}. As explained in section \ref{sec:introduction}, the current remains roughly constant over the duration of the discharge as regulated by the electronic source circuit, while the voltage and length of the arc increase continuously, the former up to a maximum of \SI{30}{V}. Data for four different current values (70, 100, 150 and \SI{250}{mA}) are used in the proceeding analyses. Images are recorded every 10 \si{\micro\second}, of which five are shown. A voltage and current value is assigned to each image as shown.

\subsection{Voltage and current}
The quasi-static relationship between voltage ($v$), current ($\iarc$) and length ($l$) in \si{\micro m} of the discharges takes the form of (\ref{eq:nott}) \cite{shekhar_modelling_2015}.
\begin{equation}
\eqalign{
v(i,l) = \vfall +\alpha l \left(1+\frac\beta{\iarc^n}\right);\quad \tz < t < \tend\cr
l(t)=\uarc t}
\label{eq:nott}
\end{equation}
The length is modelled as increasing linearly over the duration of the discharge ($\tz$ to $\tend$, as shown in figure \ref{fig:data_example} with velocity $\uarc$). The quantity $\vfall$
is known as the fall voltage, and is the minimum voltage for which the discharge can exist. Fall voltage is a property of the electrode materials, and is given a constant value of 10 V based on experimental observation. As this value compares favourably to cathode fall voltage values for cadmium given in the literature \cite{boxman1995,holm1999}, it is assumed that $\vfall$ consists entirely of a cathode fall and does not arise from anode effects. The remaining parameters $\alpha$, $\beta$ and $n$ are fitted from the data.

\begin{figure}
\includegraphics{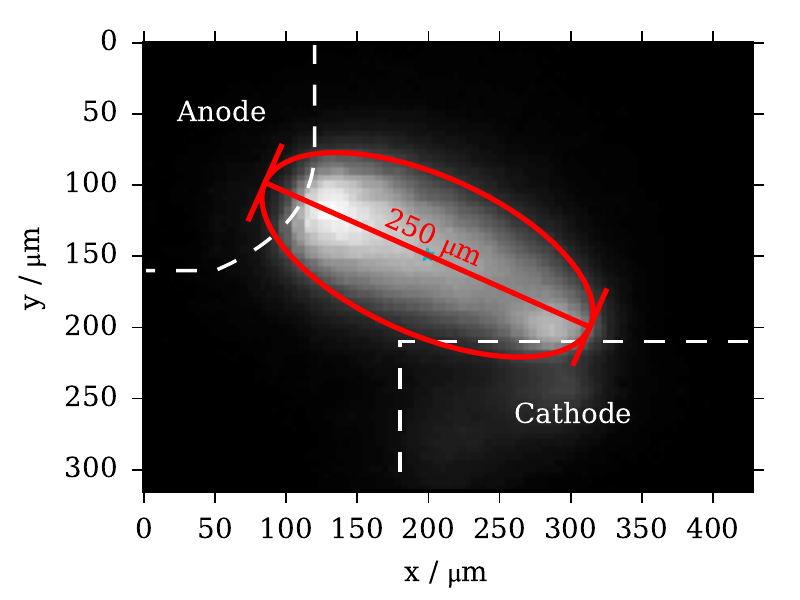}
\caption{Example of image analysis to determine arc length. Approximate locations of the electrodes are shown.}
\label{fig:arclength_example}
\end{figure}

Discharge lengths are determined using simple image analysis techniques. Here, and for the subsequent analyses of section~\ref{sec:spatial}, it is assumed for simplification that the light intensity of the discharge recorded over the spectral sensitivity of the high-speed camera (mainly the optical range) scales in good approximation with the power density of the discharge, and therefore indicates the active discharge area. Details like the change of atomic and ionic radiation intensities with species temperatures in the plasma are not considered.
In the first step of the analysis, the greyscale image of the discharge is converted to monochrome, setting to white all parts of the image with intensity greater than 20\% of the maximum value. Next, small white areas of the image are removed using a non-linear filtering operation known as morphological opening~\cite{Gonzalez2006}. The largest contiguous white region of the image is then selected, and presumed to be the discharge area. The centroid and major and minor axes of this region are then calculated, with the major axis taken as the estimate of discharge length. An example of this analysis procedure is shown in figure \ref{fig:arclength_example}. Here, the ellipse corresponding to the identified major/minor axes is shown, superimposed on the original greyscale image.

\begin{figure}
\includegraphics{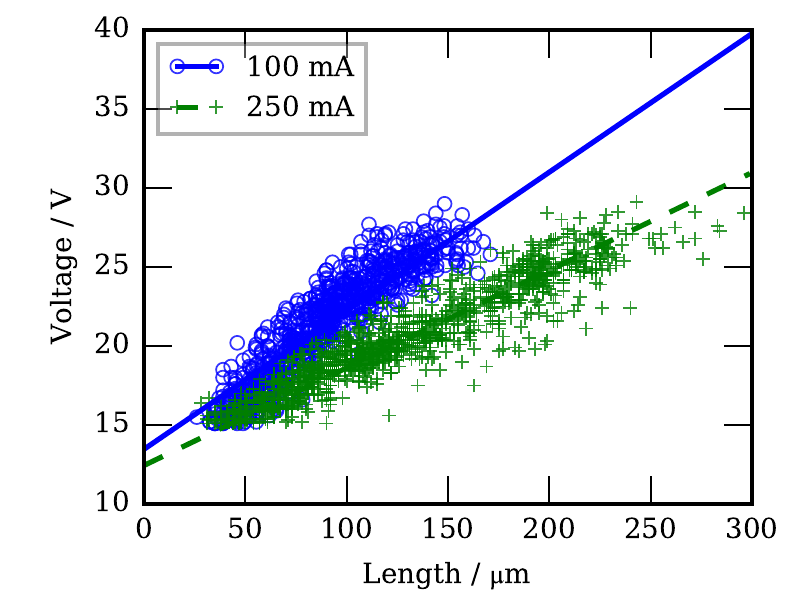}
\caption{Voltage/Length data with corresponding fitted lines from (\ref{eq:nott}). Data for two of the four current ($\iarc$) values are shown.}
\label{fig:fit_nott}
\end{figure}

The empirically determined parameters of (\ref{eq:nott}) and a systematic length error ($l_\mathrm{err}$) comprise a vector $\thv$
\begin{equation}
\thv={\qty[\alpha \quad \beta \quad n \quad l_\mathrm{err}]}^\mathrm{T}
\end{equation}
whose value can then be estimated by non-linear least squares minimisation of the function
\begin{equation}
\vb{f}(\thv)=\bm{\hat{V}}(\theta_1,\theta_2,\theta_3,\bm{I},\bm{L}+\theta_4)-\bm{V}
\label{eq:nott-errf}
\end{equation}
where $\bm{V}$, $\bm{I}$ and, $\bm{L}$ are vectors of all measured voltage data, and their corresponding currents and lengths. $\vb{\hat{V}}$ is the voltage predicted by the model (\ref{eq:nott}), and is therefore a function of the fit parameters $\theta_1$--$\theta_4$ as well as measured current and length data. Here, the systematic length error ($\theta_4=l_\mathrm{err}$) is added to the measured length. The measured data together with the fitted voltage/length relationship is shown in figure \ref{fig:fit_nott} for current values of 100 and 250 mA.

By linearising $\vb{f}$ about its least squares minimiser $\thvh$, asymptotic estimates of the standard error for each parameter $\theta_i$ can be calculated as~\cite{bates2008}
\begin{equation}
\mbox{ASE}_{\theta_i} = \frac{\left\| \vb{f}(\thvh) \right\|}{\sqrt{N-P}}
		\sqrt{\mbox{diag} \qty( \vb{J_f}(\thvh)   \vb{J_f}(\thvh)^\mathrm{T} )_i}
\end{equation}
where $N$ and $P$ are the sizes of $\vb{V}$ and $\thv$ respectively. The fitted parameter values with asymptotic estimates of the associated standard error are given in table~\ref{tab:nott_params}. The corresponding data and fitted length/voltage lines for two of the current values are shown in figure~\ref{fig:fit_nott}. The relatively large errors are a reflection of the difficulties involved in the experiment and the image based evaluation methods, as discussed in \cite{uber_experimental_2017}.

\begin{table}
\caption{\label{tab:nott_params}Fitted parameter values for (\ref{eq:nott}) with corresponding asymptotic standard error estimates (ASE)}
\lineup
\begin{indented}
\item[]\begin{tabular}{@{}lll}
\br
Parameter/Unit&Value&ASE ($\pm$\%)\\
\mr
$\alpha$ / \si{\volt \per \micro \metre}&$3.35\times10^{-2} $&$11$\\
$\beta$ / \si{\ampere \tothe n}  &$3.16\times10^{-1}$&$35$\\
$n$ / --	    &$7.09\times10^{-1}$&$\0 9$\\
\br
\end{tabular}
\end{indented}
\end{table}

\subsection{Spatial power distribution}
\label{sec:spatial}
The spatial power distribution is accounted for by another empirical relationship, namely
\begin{equation}
\eqalign{
q_{src}(r,x) = \frac{(v-\vfall)\iarc}{l\sigma^2{2\pi}}\mbox{exp}\left(- \frac{r^2}{2\sigma^2} \right)\cr
\mbox{for}\quad \tz < t < \tend, \quad 0 < x < l
}
\label{eq:qsrc}
\end{equation}
This describes the discharge as cylindrically symmetric, for axial coordinate $x$ and radial coordinate $r$. The distribution of power is thus Gaussian in the radial direction and a uniform in the axial direction. This Gaussian approximation has been used in classical models of electric arcs~\cite{mayr_beitrage_1943}. Although it is known that power density in the axial direction is likely to be non-uniform due to electrode effects, the non-uniformity cannot yet be reliably quantified from the data.

\begin{figure}
\includegraphics{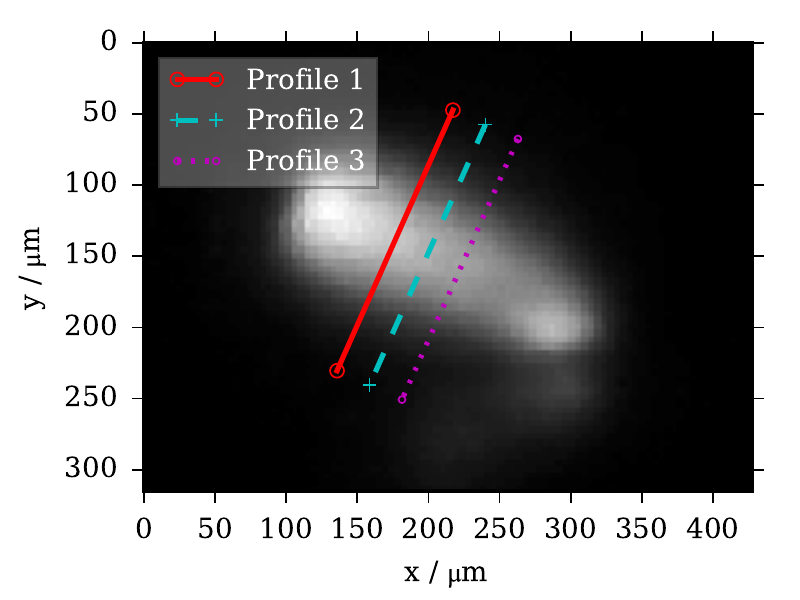}\\
\includegraphics{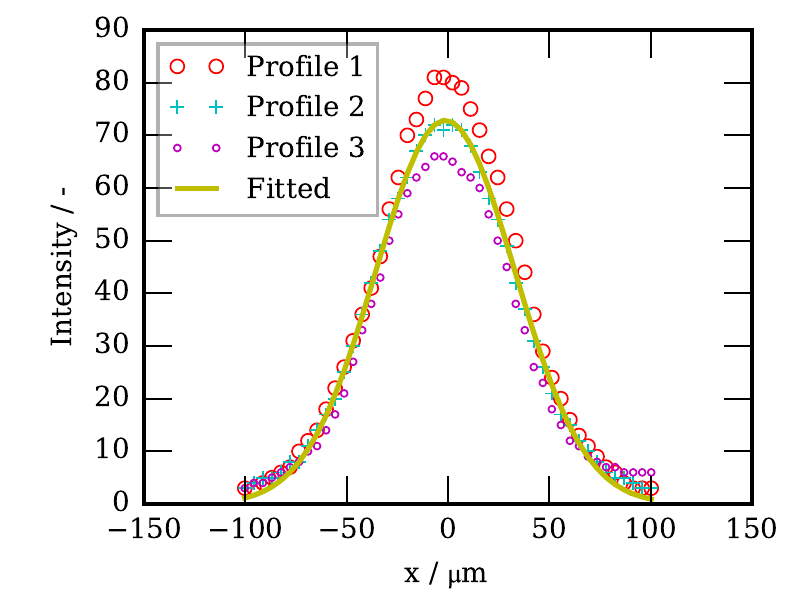}
\caption{Example of intensity line profile (above) and curve fitting (below) to determine $\sigma$ value}
\label{fig:lineprofile}
\end{figure}

\begin{figure}
\includegraphics{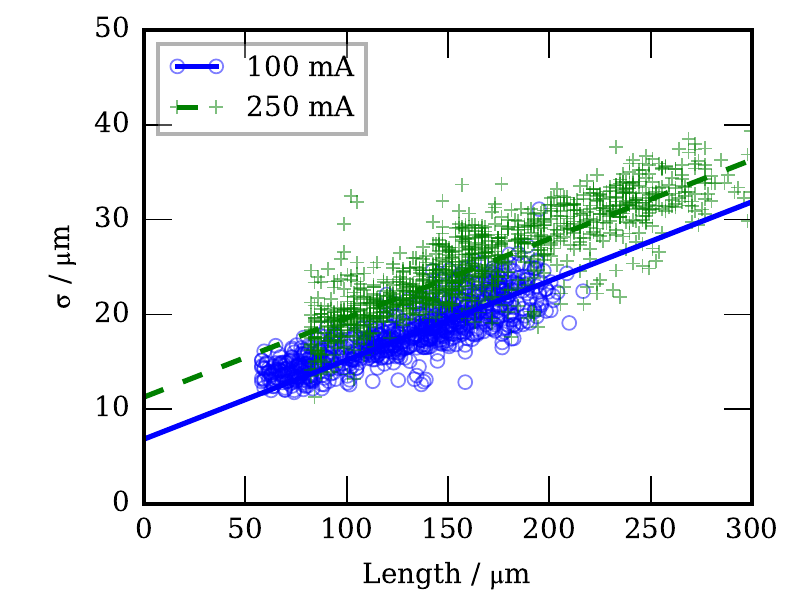}
\caption{Fitted linear function of discharge length and current for width parameter ($\sigma$). Data for two of the four current ($\iarc$) values are shown.}
\label{fig:fit_sigma}
\end{figure}

The parameter $\sigma$ of (\ref{eq:qsrc}) is derived from line profiles of intensity taken from the images (figure \ref{fig:lineprofile}). The Gaussian function is fitted to this profile data, providing an estimate of $\sigma$. This analysis is carried out across the image dataset to determine the relationship between $\sigma$ and the length and current of the discharge. In the absence of a physical model, the dependence of $\sigma$ on both quantities is presumed linear, with an expression of the form
\begin{equation}
\sigma=\sigma_0+a_\sigma l + b_\sigma \iarc
\label{eq:sigmafit}
\end{equation}
for length coefficient $a_\sigma$ and current coefficient $b_\sigma$. This expression is fitted to the data using a conventional linear least squares method, and standard errors of the parameters calculated in the usual manner \cite{bates2008}. The data and fit are show in figure \ref{fig:fit_sigma}, and parameters with corresponding standard errors of regression are shown in table \ref{tab:sigma_params}.

\begin{table}
\caption{\label{tab:sigma_params}Fitted values for parameters in (\ref{eq:sigmafit}), with standard errors of regression (SE)}
\lineup
\begin{indented}
\item[]\begin{tabular}{@{}llll}
\br
Parameter/Unit															&Value	&SE ($\pm$\%)\\
\mr
$\sigma_0$ / \si{\micro m}									&$3.88$	&$3$\\
$a_\sigma$ / --															&$0.08$	&$0.8$\\
$b_\sigma$ / \si{\micro\metre \per \ampere} &$29.5$	&$1.7$\\
\br
\end{tabular}
\end{indented}
\end{table}

\subsection{Model Validation}
\begin{figure}
  \includegraphics{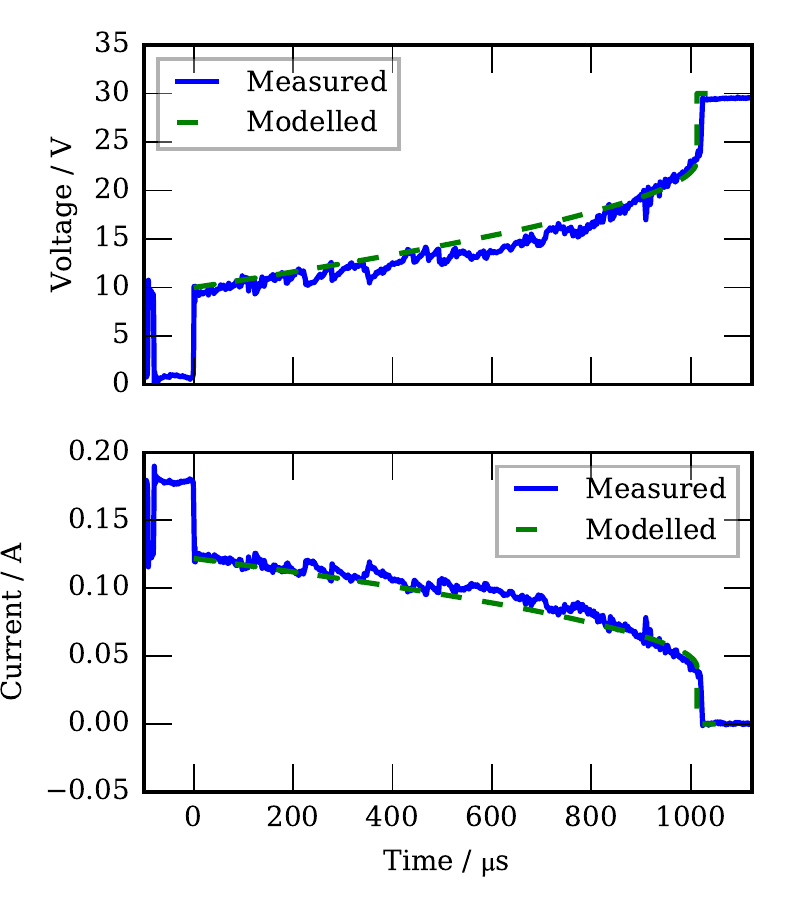}
  \caption{Comparison of model and measurement for a resistive source circuit}
  \label{fig:R_spark}
\end{figure}

\begin{figure}
  \includegraphics{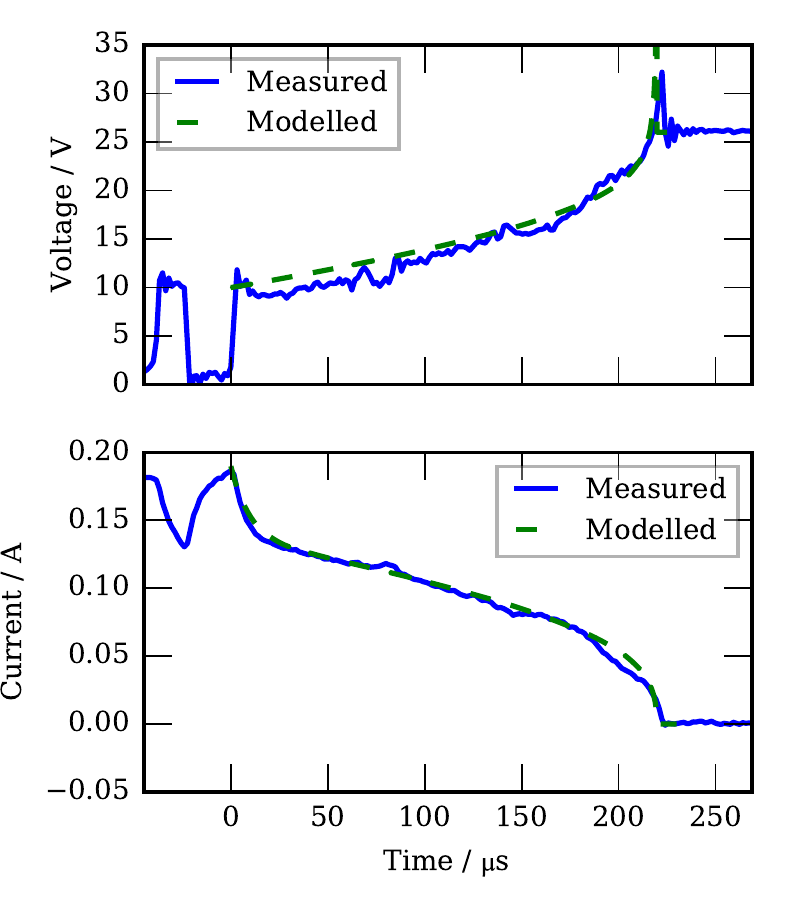}
  \caption{Comparison of model and measurement for an inductive source circuit}
  \label{fig:RL_spark}
\end{figure}

The validity of the electrical part of the empirical model can be assessed by examining voltage and current predictions under various circumstances. In section \ref{sec:introduction}, it was noted that the electrodes were connected to an electronic source circuit designed to maintain constant current. By replacing this special circuit with a simple resistive or inductive source, a discharge with time varying current is produced, and its electrical waveforms recorded. These circuits, together with a component described by (\ref{eq:nott}), can then be implemented in a common commercial circuit simulator to produce a prediction for comparison to the data. Note that only the parameter $u_\mathrm{arc}$ must be adjusted for these simulations, with the others taken from table~\ref{tab:nott_params}. A comparison of predicted and actual discharge waveforms for the resistive and inductive source circuits are shown in figures~\ref{fig:R_spark} and~\ref{fig:RL_spark} respectively.

The ability of the model to provide a good quantitative prediction under conditions outside of the original dataset used for parameter fitting is a good indication of its validity for describing the electrical characteristics of the discharge. The validity of the spatial power distribution is a more difficult question, as spatially resolved measurements of power, or even temperature, cannot be easily made. Hence, the proceeding sections attempt to shed some light on this issue by consideration of the underlying physics.

\section{Physics}
\label{sec:physics}
The empirical source term of section \ref{sec:source} is simulated in a gas dynamics model together with a simplified consideration of heat conduction and evaporation in the solid cadmium electrode via boundary conditions. A gas mixture of 21\% hydrogen in air is simulated, this being a standard mixture used to investigate ignitions by the discharges \cite{iec2011}.

\begin{figure}
\includegraphics{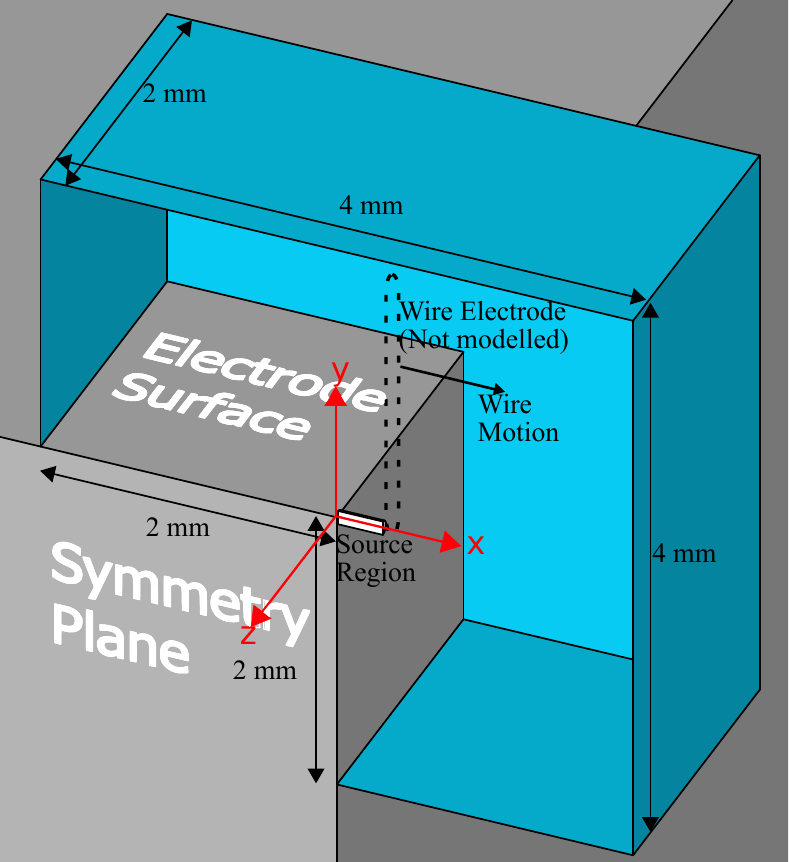}
\caption{The computational domain of the simulation. The region is bounded by a plane of symmetry (at z=0), two surfaces of the block electrode, and ``open'' boundaries (marked in blue)}
\label{fig:model_geometry}
\end{figure}

The geometry and dimensions of the computational domain are described in figure \ref{fig:model_geometry}. This is intended to represent the scenario of figure \ref{fig:contact_arc}, excluding the moving tungsten wire electrode. The relevant conservation equations and boundary conditions are discussed in this section, together with details of the calculation of electrical conductivity based on an equilibrium plasma composition.

\subsection{Conservation equations}
\label{sec:conservtion_equations}
The conservation equations of the model are given in (\ref{eq:rhoc})--(\ref{eq:Yc}). Conservation of mass is stated by,
\begin{equation}
\pdv{\rho}{t} + \div{(\rho \vb{u})} = 0 \label{eq:rhoc}
\end{equation}
for density $\rho$ and velocity $\vb{u}$. Additionally, conservation of momentum is given by the Navier-Stokes equations
\begin{equation}
\pdv{\rho \vb{u}}{t} + \div{(\rho \vb{u}\vb{u})} + \grad{p} = \div{\vb\tau} + \rho\vb{g} \label{eq:uc}
\end{equation}
for pressure $p$ and gravity vector  $\vb g = \qty[0 \quad -9.8 \quad 0]^T$. The viscous stress tensor $\vb\tau$ is calculated from Newton's law
\begin{equation}
\vb{\tau}=-\mu\qty[\grad\vb{u}+\grad{(\vb{u}^T)}-\frac23(\div\vb{u})\vb{I}]
\end{equation}
for dynamic viscosity $\mu$ and identity matrix $\vb{I}$. Conservation of enthalpy ($h$) is given by
\begin{equation}
\pdv{\rho(h+e_k)}{t} + \div{(\rho\vb{u}(h+e_k) +\vb{j_q})} - \pdv{p}{t} = q_\mathrm{src} \label{eq:hc}
\end{equation}
for heat flux $\vb{j_q}$, specific kinetic energy $e_k=\frac12 \vb{u}\vdot\vb{u}$ and $q_\mathrm{src}$ as per (\ref{eq:qsrc}). Finally, conservation of the $i$th chemical species is given by
\begin{equation}
\pdv{\rho Y_i}{t}+\div{(\rho\vb{u}Y_i + \vb{j_i})}  =  0 \label{eq:Yc}
\end{equation}
for species mass fraction $Y_i$ and diffusive flux $\vb{j_i}$. The chemical species considered are $\mbox{H}_2$, $\mbox{N}_2$, $\mbox{O}_2$ (from the gas mixture), and Cd (evaporated from the electrode). No source terms are present, as chemical reactions are not considered.

In the above equations, pressure is related to density by the ideal gas equation $p=\rho R T$ for specific gas constant $R$ and temperature $T$. Temperature is derived from enthalpy by numerically inverting the equation $h=\int_0^Tc_p \mbox{d}T$, where $c_p$ is the temperature dependent mixture specific heat, obtained from a thermodynamic database \cite{gurvich_thermodynamic_1990}. The diffusive fluxes are given by
\begin{equation}
\vb{j_i} = -\rho D_i \grad{Y_i} - \frac{D_{T_i}}{T}\grad{T} \label{eq:j_i}\\
\end{equation}
and
\begin{equation}
\vb{j_q} = -k\grad{T} + \sum h_i \vb{j_i}\label{eq:j_q}
\end{equation}
for mixture averaged diffusion coefficients $D_i$, thermodiffusion coefficients $D_{T_i}$ and thermal conductivity $k$. These transport parameters, together with viscosity $\mu$ are calculated as a function of temperature using the first Chapman-Enskog approximation~\cite{chapmen} for the pure species, and then combined using mixing rules~\cite{bird_transport_2007}.

\subsection{Boundary conditions}
The open boundaries of the computational domain are simulated using a Dirichlet boundary condition for pressure $p$=1 atm and applying Neumann boundary conditions to all other variables.

Heat conduction at the electrode surface is approximated by solving the 1-D diffusion equation in the perpendicular direction for each point on the surface
\begin{eqnarray}
\rho_s {c_p}_s\pdv{T}{t} = \pdv{q}{x_s} \label{eq:1dhc}\\
q=k_s\pdv{T}{x_s}\quad\mbox{for}\quad x_s > 0
\end{eqnarray}
where $x_s$ is the depth.  Here $\rho_s$, ${c_p}_s$, and $k_s$ and the density, specific heat and heat conductivity of cadmium, assumed constant for the simulation. The equation is solved up to a depth of 100 \si{\mu m}, corresponding to a 2 ms characteristic time of heat conduction. As suggested by \cite{giles_stability_1997},  heat flux and constant temperature boundary conditions are used for the solid (\ref{eq:1dhc}) and gas (\ref{eq:hc}) respectively, with the value for each derived from the other's solution.

The heat flux at the surface $q_{x_s=0}$ comprises of three components, namely
\begin{equation}
q_{x_s=0}  = q_\mathrm{cond} + q_\mathrm{fall} - q_\mathrm{vap}  \label{eq:qs} \\
\end{equation}
The first is the heat conducted from the gas domain.
\begin{equation}
q_\mathrm{cond} = - k\pdv{T_\mathrm{gas}}{x_s} \label{eq:qcon}\\
\end{equation}
The second represents the heating produced in the cathode fall region,  which is assumed to possess the same radially symmetric spatial distribution as the power within the discharge -- as described in (\ref{eq:qsrc})
\begin{equation}
q_\mathrm{fall} = \frac{v_\mathrm{fall}\iarc}{\sigma^2\sqrt{2\pi}}
	\mbox{exp}\left(- \frac{r_s^2}{2\sigma^2} \right) \label{eq:qfall}
\end{equation}
where $r_s$ represents distance along the surface from the origin of the discharge. It is also assumed here that this energy is completely transferred towards the cathode surface and not to the plasma. The last component  represents heat lost due to evaporation
\begin{equation}
q_\mathrm{vap}  =	(L_v+{c_p}_s T_{x_s=0})J_m \label{eq:qvap}
\end{equation}
where $L_v$, is the specific heat of vaporisation. $J_m$ is the surface normal mass flux  due to evaporation. This is calculated with
\begin{equation}
J_m = -\rho D\rs{Cd} \dv{Y\rs{Cd}}{x_s} + \alpha_e\frac{p_v}{\sqrt{2 \pi R\rs{Cd}T_{x_s=0}}} \label{eq:Jm}
\end{equation}
where the first term represents diffusion flux and the second bulk evaporation according to the Langmuir relation~\cite{benilov2001}. Here, $D\rs{Cd}$, $Y\rs{Cd}$ and $p\rs{Cd}$ represent the diffusion coefficient, mass fraction and partial pressure of cadmium at the surface. $R\rs{Cd}$ is the specific gas constant for cadmium. An accommodation coefficient $\alpha_e$ is calculated according to the procedure of \cite{benilov2001}. The equilibrium vapour pressure $p_v$ is estimated as a function of temperature from the August equation coefficients given in \cite{alcock_vapour_1984}.

To account for the evaporation process, Dirichlet boundary conditions on the surface are also imposed for velocity in the surface normal direction ($|\vb{u}| = J_m/\rho$) and cadmium mass fraction ($Y\rs{Cd} = p_v M\rs{Cd}/(pM)$), where $M\rs{Cd}$ and $M$ are the molar mass of cadmium and and the mixture average molar mass respectively.

\subsection{Plasma}
\label{sec:plasma}
For the calculation of electrical conductivity, a pure cadmium plasma at a pressure equal to the partial pressure of cadmium in the gas mixture has been assumed. This assumption is based on the previous results showing the discharge radiation to comprise solely of cadmium lines, as mentioned in section \ref{sec:introduction}. Determination of the transport properties of this cadmium plasma has been performed using the procedure presented in~\cite{poritsky2013}. First, the plasma composition has to be calculated. Under the assumption of local thermodynamic equilibrium (LTE), the temperature dependent densities of plasma species, namely, electrons, atoms and ions, can be predicted using the method of the minimisation of generalised Gibbs free energy~\cite{poritsky2013}. The following components have been taken into account: electrons $e$, atoms Cd, ions Cd$^+$, Cd$^{2+}$, Cd$^{3+}$, Cd$^{4+}$ and Cd$^{5+}$. The densities have been calculated for various vapour pressures. When the densities are known, transport parameters for the corresponding pressures and temperatures can be calculated.

\begin{figure}
\includegraphics{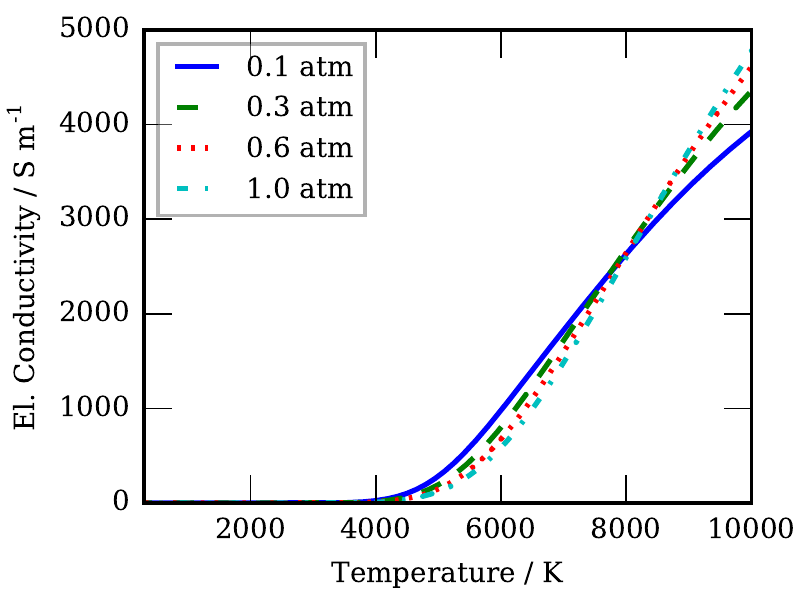}
\caption{Electrical conductivity of Cd plasma at different vapour pressures}
\label{fig:Cdelcond}
\end{figure}

Similarly to the calculation of gas transport parameters (section~\ref{sec:conservtion_equations}), the Chapman-Enskog theory is utilised for obtaining transport properties of the thermal plasma in equilibrium state~\cite{chapmen}. In the specific case of a monatomic gas, transport parameters can be obtained using the approach developed by Devoto~\cite{devoto}. The starting point for the determination of transport coefficient is the calculation of kinetic integrals which requires the knowledge of the scattering cross sections as input data. Kinetic integrals for Cd-Cd collisions have been calculated using the Lennard--Jones model potential with parameters from~\cite{rabani}. For the determination of the data for the collisions between electrons and Cd atoms scattering cross sections from \cite{kontrosi, nahar} have been adopted. In order to describe the ion-atom interaction the common assumption that the interaction cross section for the ion-atom pair is generally defined by polarisation ion capture by the atom has been used.
The polarisability of cadmium  has been taken from~\cite{CRC90}. Finally, interaction between charged particles has been characterised by the Coulomb cross-section~\cite{liboff1959}. Figure~\ref{fig:Cdelcond} shows the resulting temperature dependence of electrical conductivity for different vapour pressures.

\section{Simulation Results}
\label{sec:results}
The equations of section~\ref{sec:physics} are implemented in the OpenFOAM finite volume method solver \cite{weller1998}. The pressure implicit with splitting of operator (PISO) method is used to solve for pressure and velocity \cite{issa1986}. Transport parameters in the conservation equations are calculated using the Cantera software \cite{goodwin2016}.

\begin{figure}
\includegraphics{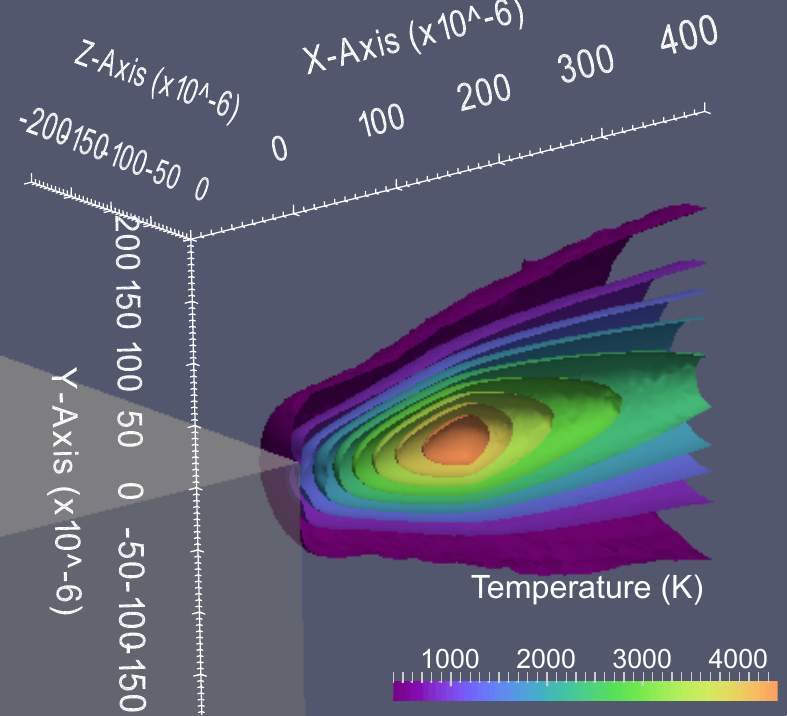}
\includegraphics{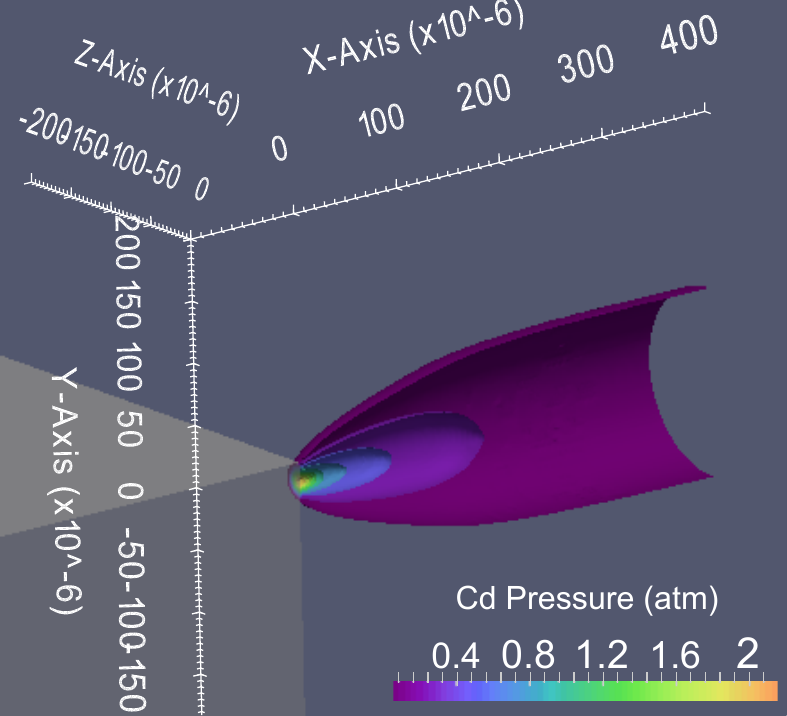}
\includegraphics{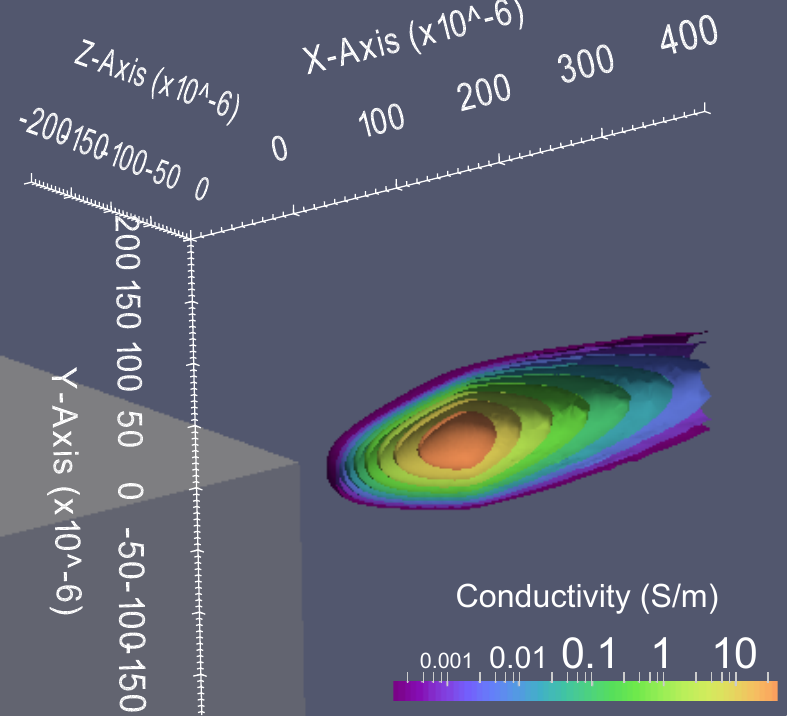}
\caption{Example isosurface plots for subset of field data for (from top) temperature, cadmium partial pressure and electrical conductivity. Data is from case 2 at t= 150 \si{\micro s}.}
\label{fig:field_data_ex}
\end{figure}

Simulations are conducted for the discharge parameter values given in table \ref{tab:sim_cases}, these being representative of the data collected in \cite{uber_experimental2_2017} and of the application in general. Here, the simulated discharges are specified in terms of current ($\iarc$) and duration ($\tend$), from which velocity $\uarc$ is calculated using (\ref{eq:nott}). For each simulation, the time varying fields of mass fraction for cadmium are used to calculate the corresponding partial pressure. The partial pressure and temperature fields are then used to calculate an electrical conductivity field using the data of section \ref{sec:plasma}. Example fields for temperature, cadmium partial pressure and electrical conductivity are shown in figure \ref{fig:field_data_ex}.

\begin{table}
\caption{\label{tab:sim_cases} Overview of simulated cases. Parameter values are as described in figure \ref{fig:data_example} and (\ref{eq:nott})}
\lineup
\begin{indented}
\item[]\begin{tabular}{@{}llll}
\br
Case		&Current / mA	&Duration / \si{\micro s}	&$\uarc$ / \si{\metre \per \second}\\
\mr
1			&70				&200					&0.96\\
2			&100			&200					&1.14\\
3			&70				&1000					&0.19\\
4			&100			&1000					&0.23\\

\br
\end{tabular}
\end{indented}
\end{table}

\begin{figure}
	\includegraphics{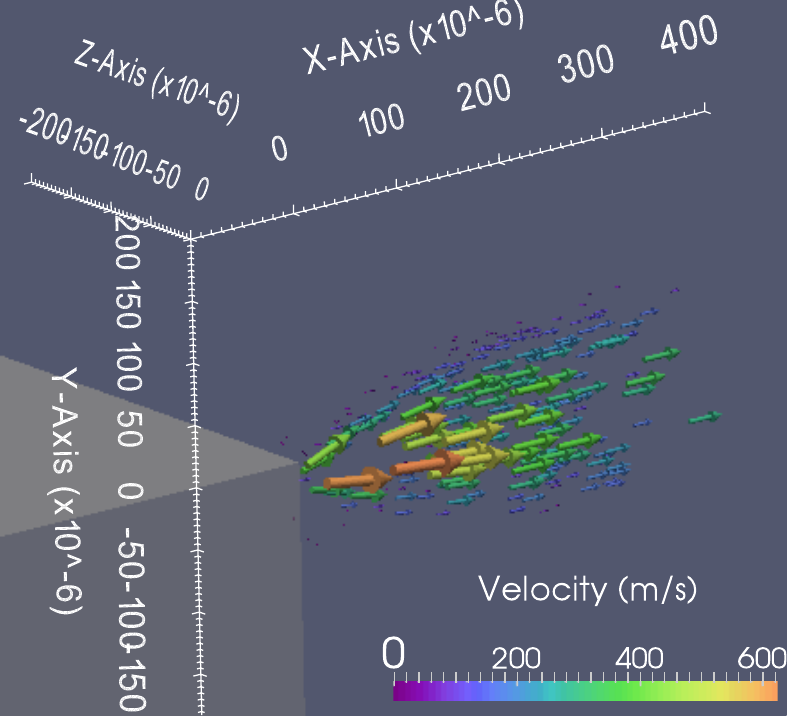}
	\caption{Velocity plot from case 2 at t=150 \si{\micro s}}
	\label{fig:U_data_ex}
\end{figure}

A notable characteristic of figure \ref{fig:field_data_ex} is the elongated distribution of temperature and cadmium partial pressure. Note that the discharge length at this point is around 170 \si{\micro m}, however the high temperature region extends well beyond this length. This is due to a jet of evaporated material produced by the cathode fall region heating, which can be seen in the velocity distribution of figure \ref{fig:U_data_ex}.

\begin{figure}
\includegraphics{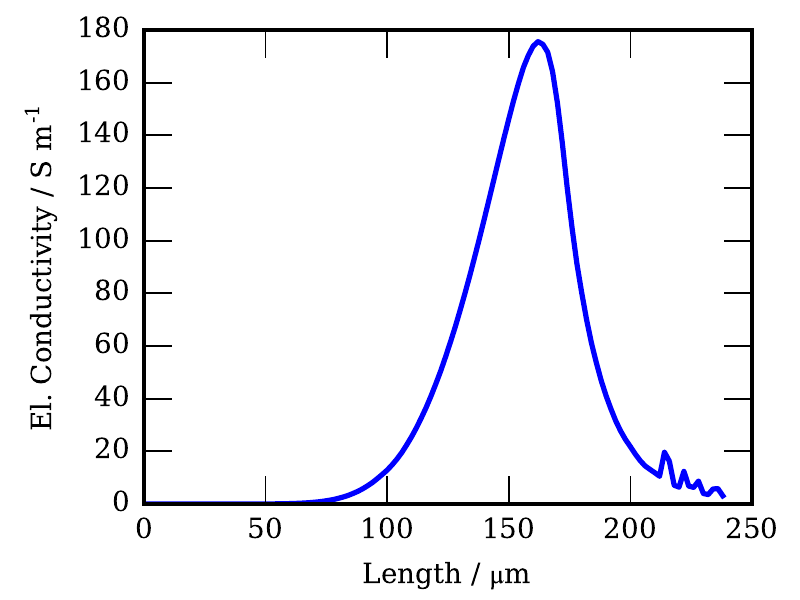}
\includegraphics{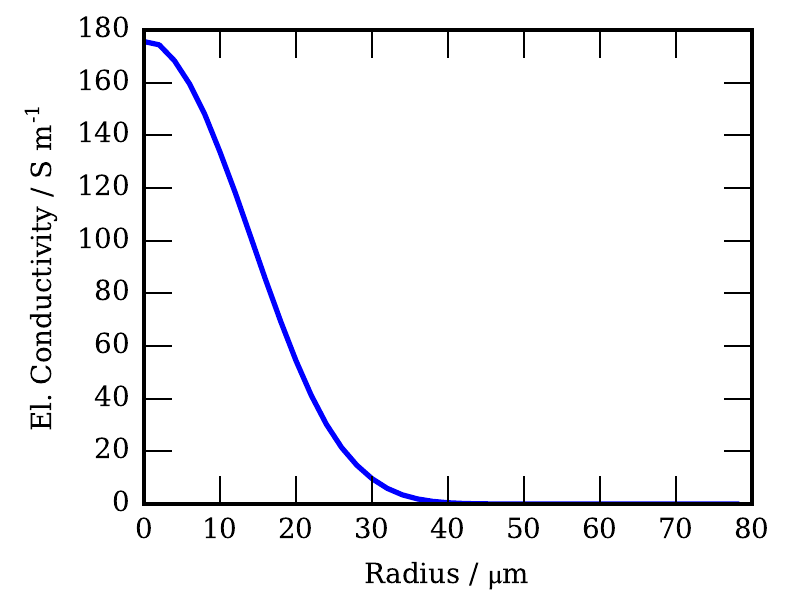}%
\caption{Example of axial (at r=0) and average radial (at l=168 \si{\micro m}) profile of electrical conductivity, for simulated case 2 at t= 150 \si{\micro s}.}
\label{fig:sigma_profile}
\end{figure}

The electrical conductivity field is then sampled to produce an averaged distribution in 2-D cylindrical coordinates. An example of the axial and average radial profile of electrical conductivity is shown in figure \ref{fig:sigma_profile}.

The sampled axisymmetric distribution of electrical conductivity can be numerically integrated to estimate overall resistance $Z$.
\begin{equation}
Z = \int\limits_{L_0}^{L_1}\frac{dl}{\int_{0}^{R_1} 2\pi r \sigma_e dr} \label{eq:resistance}
\end{equation}
The radial limit of integration $R_1$ can simply be made large enough to include the conductive region (200 \si{\micro m} is selected here). The axial limits $L_1$ and $L_0$ require consideration. As the axial profile of figure \ref{fig:sigma_profile} shows, conductivity is negligible near the ends of the discharge, particularly on the cathode (left) side. Since electrode layer effects are not modelled, including the areas near the electrodes in the integration would not be meaningful. The analysis is thus conducted by integrating only over the region where conductance is no less than 20\% of its maximum value. Under these conditions, the evaluation of (\ref{eq:resistance}) will provide a lower bound on possible resistance.

\begin{figure}
\includegraphics{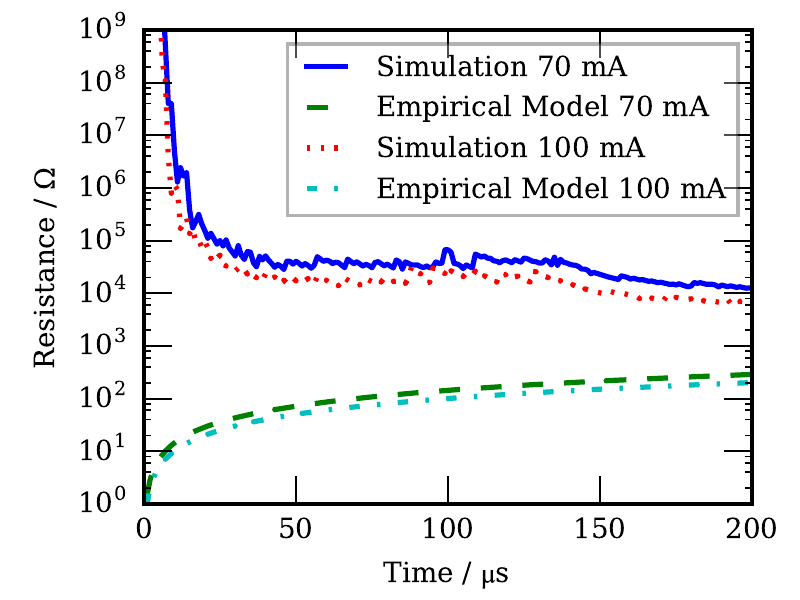}%
\caption{Comparison of discharge resistance variation over time, for simulated discharges of 200 \si{\micro m} (cases 1 and 2)}
\label{fig:res_comp_200us}
\end{figure}

\begin{figure}
\includegraphics{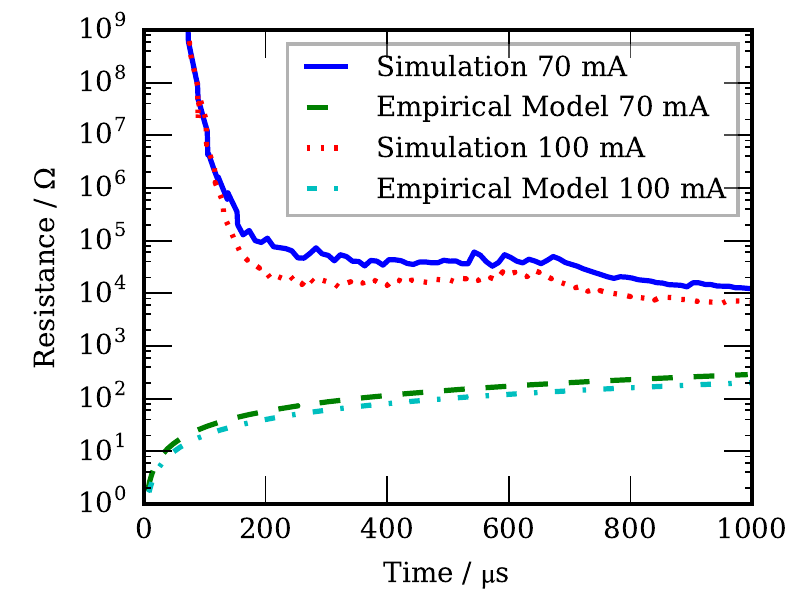}%
\caption{Comparison of discharge resistance variation over time, for simulated discharges of 1000 \si{\micro m} (cases 3 and 4)}
\label{fig:res_comp_1000us}
\end{figure}

The resistance at each time step of the simulated cases was calculated both (\ref{eq:resistance}) and the empirical relationships of (\ref{eq:nott}), as $Z=v(i,l)/\iarc$. A comparison of the results from both calculation methods can be seen in figures \ref{fig:res_comp_200us} and \ref{fig:res_comp_1000us}, for the simulated discharges of 200 and 1000 \si{\micro s} respectively. Here, the expected values are those predicted by (\ref{eq:nott}) as described above. The comparison provides an indication of consistency of the model under the LTE assumption. The results show an inconsistency of at least an order of magnitude, which is particularly stark at the beginning of the discharge period.

\section{Discussion}
\label{sec:discussion}
The results show that an LTE temperature--electrical conductivity relationship  cannot adequately describe the discharge in question. Even under the favourable conditions assumed here, (i.e.: no radiation losses, near electrode regions assumed perfectly conducting), the discrepancy between electrical resistance predicted this relationship, and that predicted by an empirical correlation is not less than one order of magnitude. In addition to the difference in magnitude of the electrical resistance, the way in which this resistance varies over the duration of the discharge is also noteworthy. Under the LTE assumption, the resistance would be expected to decrease over time. This is because the average temperature increases over time, and the electrical resistance decreases near-exponentially with temperature (and increases only linearly with length). The data and empirical correlations of section~\ref{sec:source}, however, show that voltage is linearly proportional to length. This implies a constant electrical resistance per unit length over the duration of the discharge.

From these observations, it can be reasonably concluded that the properties of the plasma are not (or at best, very weakly) dependent on the temperature of the gas. This is a defining characteristic of a non-equilibrium (or non-thermal) plasma. Although further investigation is required to determine the exact nature of the plasma, it could possibly be described as a type of glow discharge. An obvious question arising from this hypothesis is how the empirical model of section~\ref{sec:source} is able to describe the electrical characteristics of the discharge with reasonable accuracy. This is surprising, given that the hyperbolic relationship between voltage and current is based on LTE, specifically, the idea that higher currents produce a broader and hotter (therefore more conductive) discharge column. It is possible that the apparent applicability of the model is coincidental. One theory of stable glow discharges suggests that a constant current density at the cathode is maintained, with increases in current only increasing the size of the electron emitting area (or cathode spot)~\cite{raizer1991}. It is conceivable that this would, in turn, also increase the width of the discharge column, leading to a qualitatively similar voltage/current relationship as in the case of a stable arc. It should be stressed that these ideas are speculative, and require confirmation through experiment and/or more detailed simulations.

\section{Conclusion}
\label{sec:conclusion}
The results have significant implications for the simulation of combustion initiated by the electrical discharges in question. Thus far, the empirical modelling approach has shown some success in predicting electrical properties of the discharge. In the prediction of flammable gas ignition, however, the gas temperature, or more accurately, the heavy particle translational temperature, is also an important parameter. The current empirical approach detailed in sections \ref{sec:introduction} and \ref{sec:source} assumes that all of the electrically measured power is thermalised, and distributed spatially in a similar manner to the discharge's observed radiation intensity. This is problematic, firstly because the power loss due to this radiation is disregarded, and secondly, because the distribution is difficult to accurately measure by experiment. Although previous research~\cite{shekhar_ignition_2017} found that qualitative features of the flammable gas ignition were adequately predicted by this empirical approach, a more physically realistic approach would be more satisfactory, and could potentially approach quantitative accuracy without arbitrary model tuning. Conventional simulation methods which use a temperature--electrical conductivity curve to couple hydrodynamics to electromagnetic equations cannot be used to solve this problem, as they rely on the LTE assumption.

An improvement on the experimental side would require more sophisticated measurement apparatus, such as absolutely calibrated spectroscopy or spectrally resolved imaging. On the numerical side, improving the physical correctness of the model would only be possible by accounting for the non-equilibrium thermodynamics and associated radiative processes. A more rigorous treatment of the electrodes, especially electron emission, is also required. These extensions to the model pose significant challenges due to the uncommon material (cadmium) involved, for which limited published data exists.

\section*{Acknowledgements}
The authors wish to thank Carsten Uber and Steffen Pohl for the provision of the experimental dataset, Mohammed Abdul Moiz for the creation of a program to automate the image based measurements, and Thomas Uehlken for the development of the experimental apparatus.

\section*{References}
\bibliography{shekhar,cd}
\bibliographystyle{iopart-num}

\end{document}